\documentstyle [twocolumn,epsf]{mn}
\newcommand{\mincir}{\raise
-3.truept\hbox{\rlap{\hbox{$\sim$}}\raise4.truept\hbox{$<$}\ }}
\newcommand{\magcir}{\raise
-3.truept\hbox{\rlap{\hbox{$\sim$}}\raise4.truept\hbox{$>$}\ }}
\newcommand{\minmag}{\raise
-3.truept\hbox{\rlap{\hbox{$<$}}\raise5.truept\hbox{$<$}\ }}
\newcommand{\be}{\begin{equation}}
\newcommand{\ee}{\end{equation}}
 \newcommand{\ba}{\begin{eqnarray}}
\newcommand{\ea}{\end{eqnarray}}
\newcommand{\brr}{\begin{array}}
\newcommand{\err}{\end{array}}
\newcommand{\bc}{\begin{center}}
\newcommand{\ec}{\end{center}}

\newcommand{\lb}{{\left<\right.}}
\newcommand{\rb}{{\left.\right>}}

\title[Large scale structure in the HIPASS Survey]
{Large scale structure in the HI Parkes All-Sky Survey: Filling the
  Voids with HI galaxies?}
\author[Basilakos et al.]{S. Basilakos$^{1}$, M. Plionis$^{2,3}$, 
K. Kova\v{c}$^{4}$, and N. Voglis$^{1, \dag}$ \\
\vspace{0.1cm}
$^{1}$ Academy of Athens, Research Center for Astronomy \& Applied
  Mathematics, Soranou Efessiou 4, 11-527, Athens, Greece\\
$^2$ Institute of Astronomy \& Astrophysics, National Observatory of Athens, 
I.Metaxa \& B.Pavlou, Palaia Penteli, 152 36, Athens, Greece \\
$^3$ Instituto Nacional de Astrof\'{\i}sica, \'Optica y Electronica (INAOE)
Apartado Postal 51 y 216, 72000, Puebla, Pue., M\'exico\\
$^{4}$Kapteyn Astronomical Institute, University
of Groningen, The Netherlands \\
$\dag$ deceased 9/2/2007.
}

\begin{document}

\maketitle

\begin{abstract}
We estimate the two-point correlation function in redshift space
of the recently compiled HIPASS neutral hydrogen (HI) sources 
catalogue, which if modeled as a power law, $\xi(r)=(r_{0}/r)^{\gamma}$,  
the best-fitting parameters for the HI selected galaxies are found to be 
$r_{0}=3.3 \pm 0.3 \; h^{-1}$ Mpc with 
$\gamma=1.38 \pm 0.24$.
Fixing the slope to its universal value $\gamma=1.8$, we obtain
$r_{0}= 3.2\pm 0.2 \; h^{-1}$ Mpc.

Comparing the measured two point correlation function 
with the predictions of the concordance cosmological model
($\Omega_{\Lambda}=0.74$), 
we find that at the present epoch the HI selected galaxies are
anti-biased with respect to the underlying matter fluctuation 
field with their bias value being $b_{0}\simeq 0.68$. Furthermore,
dividing the HI galaxies into two richness subsamples we find that 
the low mass HI galaxies have a very low present bias factor
($b_{0}\simeq 0.48$), while
the high mass HI galaxies trace the underlying matter distribution
as the optical galaxies ($b_{0}\simeq 1$). 
Using our derived present-day HI galaxy bias we estimate their
redshift space distortion 
parameter, 
and correct accordingly the correlation function for peculiar motions.
The resulting real-space correlation length is
$r^{\rm re}_{0}=1.8 \pm 0.2 \;h^{-1}$Mpc and 
$r^{\rm re}_{0}=3.9 \pm 0.6 \;h^{-1}$Mpc for the low and
high mass HI galaxies, respectively. The low-mass HI galaxies appear to have
the lowest correlation length among all extragalactic populations
studied to-date. In order to corroborate these results we have
correlated the IRAS-PSCz reconstructed density field, 
smoothed over scales of 5
$h^{-1}$ Mpc, with the positions of the HI
galaxies, to find that indeed the HI galaxies are typically 
found in negative overdensity regions
($\delta\rho/\rho_{\rm PSCz} \mincir 0$), 
even more so the low HI-mass galaxies.

Finally, we also study the redshift evolution of the HI 
galaxy linear bias factor and find that the HI-galaxy population 
is anti-biased up to $z\sim 1.3$. While at large redshifts
$z \sim 3$, we predict that the HI galaxies are strongly biased. 
Our bias evolution predictions are
consistent with the observational bias results of Lyman-$\alpha$ galaxies. 

\end{abstract}

\vspace{1.0cm}

\section{Introduction}
The clustering 
of the different extragalactic sources is an ideal tool for 
testing theories of structure formation as well as studying the 
large-scale structure (eg. Peebles 1993).
The traditional indicator of clustering, the 
two-point correlation function, is a fundamental statistical test
for the study of the extragalactic distribution and is relatively
straightforward to measure from observational data. 
Recently the new generation redshift surveys such as the 
Point Source Catalogue for Redshift (PSC-z; Saunders et al. 2000), 
the European Large-Area {\it ISO} survey (ELAIS; Oliver et al. 
2000), the Sloan Digital Sky Survey (SDSS; York et al. 2000)
the 2dF Galaxy Redshift Survey (2dFGRS; Colless et al. 2001),
the DEEP2 Galaxy Redshift Survey (DEEP2; Davis et al. 2003)
and the VIMOS-VLT Deep Survey (VVDS; 
Le F\'{e}vre, et al. 2005)
have extended the clustering studies and now we are able to 
extract definitive measurements of galaxy clustering in the 
local and relatively distant Universe.

Indeed Zehavi et al. (2002) and Hawkins et al. (2003) have found 
$r_{0}\simeq 5 \;h^{-1}$Mpc and $\gamma\simeq 1.7$ for the SDSS
galaxies, while Norberg et al. (2001) found for the 2dFGRS galaxies that
$r_{0}\simeq 4.9h^{-1}$Mpc and $\gamma\simeq 1.7$. 
Also, Coil et al. (2006) have found 
$r_{0}\simeq 4 \;h^{-1}$Mpc and $\gamma\simeq 1.7$ for the DEEP2
galaxies at $z\sim 1$ and Guzzo et al. (2007) derived 
$r_{0}\simeq 3.6 \;h^{-1}$Mpc for the VVDS
galaxies at $z\sim 1.5$.
In the infrared side,
Jing, B\"orner \& Suto (2002) derived 
$r_{0}\simeq 3.7h^{-1}$Mpc and $\gamma\simeq 1.7$, using the 
IRAS-PSCz data, while
Gonzalez-Solares et al. (2004) using the ELAIS sources found
$r_{0}\simeq 4.3h^{-1}$Mpc and $\gamma\simeq 2$
correlation functions.
From a cosmological point of view, it is imperative to understand
how the different types of galaxies trace the
underlying mass distribution. It is well known that the
large scale clustering pattern of different mass tracers 
(galaxies, AGN, clusters, etc) is characterized by a bias picture 
(Kaiser 1984; Bardeen et al. 1986). 
In particular, biasing is assumed to be statistical 
in nature; galaxies and clusters are identified as high peaks
of an underlying, initially Gaussian, random density field.
Biasing of galaxies with respect to the dark matter distribution was
also found to be an essential ingredient of Cold Dark Matter (CDM) 
models of galaxy formation in order to reproduce the observed galaxy 
distribution (eg. Benson et al. 2000).  

In this paper we utilize the recently completed HI Parkes All-Sky 
Survey (HIPASS; Barnes et al. 2001), which is the 
largest uniform sample of 
HI selected galaxies in the local Universe, 
attempting to make a detailed investigation of the 
connection between the clustering and the biasing properties of 
HI selected galaxies.
In particular we determine, in the framework of the concordance $\Lambda$CDM 
cosmology, the relative HI galaxy bias at the 
present time and describe the corresponding bias as a function of redshift.
For a detailed study of the observational 
clustering properties of the HIPASS sources we refer 
the reader to the recent works of Ryan-Weber (2006) and Meyer
et al. (2007). We also refer the reader to the new
{\em Arecibo Legacy Fast ALFA} ({\sc alfalfa}) drift survey, especially
designed to probe the faint end of the HI mass function in the local
universe (Giovanelli et al. 2005a; 2005b).

The structure of the paper is as follows. In section 2 we discuss
the HI galaxy dataset, its measured redshift-space correlation
function, the determination of its present day bias factor, the
corrected for redshift space distortion real-space correlation 
function as well as an investigation of the environment of HI galaxies.
In section 3 we present two bias evolution models and study the
corresponding evolution for the case of HI galaxies within 
the concordance $\Lambda$CDM model. 
Finally, we draw our conclusions in section 4.

\section{Estimation of the HI selected galaxies correlation function}
\subsection{The HIPASS data}
The recent HIPASS catalogue
(for more details see Meyer et al. 2004; Ryan-Weber 2006), 
contains 4315 neutral hydrogen (HI) 
sources in the southern region of the survey ($\delta<+2^{\circ}$),
covering an area of $\sim 2.07\pi$ on the sky between
300 km s$^{-1}$ and 12700 km s$^{-1}$ 
(see also Doyle et al. 2005). 
We exclude detections
of which the expected completeness is very low ($C<0.5$) in order 
to avoid introducing systematic effects related to the HI galaxy
space density (estimation of the completeness limit $C$ is given 
in Zwaan et al. 2004; 2005). The 
remaining sample contains $\sim 4008$ HI selected galaxies with 
masses $M_{\rm HI}\le 4\times 10^{10} h^{-2}M_{\odot}$ 
\footnote{Zwaan et al. (2005) used 
$H_{0}=75h_{75}\,$km$\,$s$^{-1}\,$Mpc$^{-1}$. In that 
case the maximum HI mass is 
$\sim 7.1\times 10^{10} h_{75}^{-2}M_{\odot}$}.
In the mean time northern extension of the HIPASS has been
published containing 1002 HI detections in the following area 
$2^{\circ}<\delta<20^{\circ}$ (Wong et al. 2006).

The measured radial velocities are translated to the Local group frame
and are converted to proper distances using a spatially flat cosmology with 
$H_{0}=100 \;h\,$km$\,$s$^{-1}\,$Mpc$^{-1}$ and 
$\Omega_{\rm m}=1-\Omega_{\Lambda}=0.26$.

\begin{figure}
\mbox{\epsfxsize=11cm \epsffile{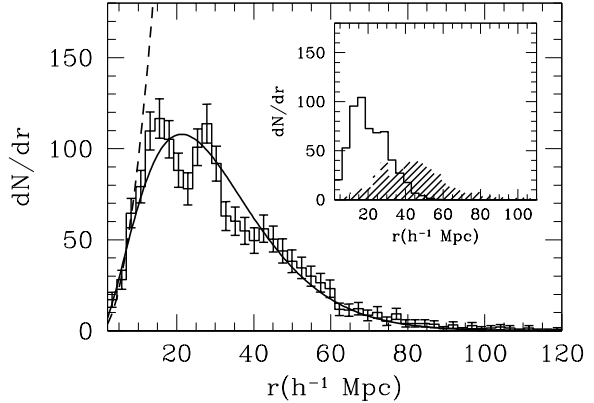}}
\caption{The measured (histogram) and the fitted 
(continuous line) number of the
HI galaxy sources as a function of distance. The dashed line corresponds 
to a volume limited expectation. The insert panel
shows the corresponding selection function for the $S_{1}$ (solid)
and the $S_{2}$ (hatched) subsamples, respectively (see section 2.2).} 
\end{figure}

In Fig. 1, we present the estimated (histogram) 
and that expected for a volume limited sample (dashed line), number 
of the HI sources as a function of distance. It appears that the
HI galaxy distribution is roughly volume-limited 
out to $10 h^{-1}$Mpc, a fact corroborated also by the $\chi^{2}$ 
minimization test between model and observations,
which gives a reduced 
$\chi^{2}/{\rm dof}$ (up to $r \le 10h^{-1}$Mpc) of $1.04$.
However, due to the fact that the HIPASS catalogue is a peak flux 
limited sample it suffers from the well known degradation of sampling as 
a function of distance (codified by the so called {\it selection function}). 
Thus, we can attempt to parametrize the HI galaxy distance distribution,
for this particular effect, using the following formula:
\be 
\frac{ {\rm d}N}{{\rm d}r}= \rm {d}\Omega r^{2} \lb \rho \rb \phi(r) 
\ee
where $\phi(r)$ is the parametrized selection function
\be
\phi(r)=\frac{A}{\rm{d}\Omega \lb \rho \rb}
r^{\alpha} {\rm exp}\left[ -(r/r_{s})^{m} 
\right]
\ee
with 
\be
A=\frac{N m}{\Gamma[(\alpha+3)/m] \; r_{s}^{\alpha+3}} \;\;.
\ee
Note that $N$ is the total number of the HI sources, ${\rm d}\Omega$ 
is the solid angle of the survey, 
$\lb \rho \rb$ is the mean density of the HIPASS sample (see below)
and the factor $\Gamma$ 
is the Gamma function.
The solid line in Fig.1 corresponds to the best-fitting 
${\rm d N}/{\rm d}r$, which is determined 
by the standard $\chi^{2}$ minimization procedure in which each
part of the histogram is weighted by its Poisson error.
The reduced $\chi^{2}/{\rm dof}$ is $\sim 0.90$ and the corresponding values
of the fitted parameters are $r_{s}=16.5^{+1.0}_{-2.5} h^{-1}$Mpc, 
$m=1.24^{+0.06}_{-0.08}$
and $\alpha=-0.30 \pm 0.12$ (the uncertainties correspond to 
1$\sigma$ errors).  
In Fig. 2 we present the number density of our sample as a 
function of distance. Up to $\sim 10 \;h^{-1}$ Mpc the 
corresponding space density is roughly constant, being 
$\lb \rho \rb=0.14\pm 0.02 \;h^{3}\,$Mpc$^{-3}$ 
and then drops because of the sample flux limit.

\begin{figure}
\mbox{\epsfxsize=9.6cm \epsffile{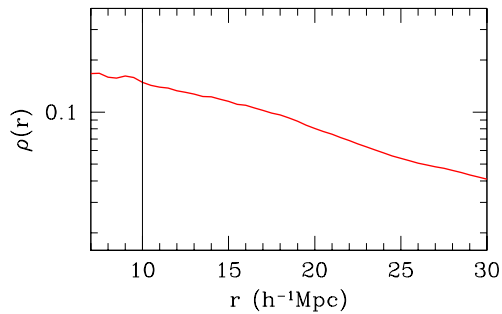}}
\caption{The space density of the HIPASS galaxy data as a function 
of distance.} 
\end{figure}

\subsection{HI selected galaxy redshift-space correlation function}
We estimate the redshift space HI galaxy correlation function 
using the estimator described by Efstathiou et al. (1991):

\be
\xi_{z}(r)=f\frac{N_{DD}}{N_{DR}}-1,
\ee 
where $N_{DD}$ is the number of 
HI pairs in the interval $[r-\Delta r,r+\Delta r]$, and
$N_{DR}$ is the number of data-random pairs. 
In the above relation the normalization factor is $f = 2 N_R /(N_D-1)$, with
$N_D$ and $N_R$ the total number of data and random points,
respectively. 
The random distributions were constructed 
using 100 Monte Carlo realizations, respecting the survey area and
taking into account all possible 
systematic effects of the data (eg. completeness, fraction of HI 
sources missed by the finding algorithm due to the HIPASS flux limit
and the redshift distribution of the HI galaxy data).
Note, that we have tested that our random catalogues reproduce exactly 
the corresponding HI selection function. 

\begin{table*}
\caption[]{Results of the correlation function analysis 
of the HIPASS sample.
Errors of the fitted parameters represent $3\sigma$ uncertainties.
Finally, the $r_{0}$ has units of $h^{-1}\,$Mpc.}

\tabcolsep 6pt
\begin{tabular}{cccccccccc} 
\hline
Sample & $N$& $r_{0}$ & $\gamma$ & $r_{0}(\gamma=1.8)$ & $b_{0}$ & 
$\beta$ & $K(\beta)$ & $r^{\rm re}_{0}$ &$r^{\rm re}_{0}(\gamma=1.8)$\\ \hline \hline 
All& 4008&$3.3\pm 0.3$ & 1.38$\pm 0.24$& $3.2\pm 0.2$&$0.68\pm 0.10$ & 0.66&1.52&
$2.4\pm 0.3$& $2.5\pm 0.2$\\
$S_{1}$& 2389&$2.4\pm 0.3$ & 1.72$\pm 0.46$& $2.4\pm 0.2$&$0.48\pm 0.10$ & 0.93&1.79&
$1.8\pm 0.3$& $1.8\pm 0.2$\\
$S_{2}$& 1619&$4.9\pm 1.0$ & 1.40$\pm 0.42$& $4.6\pm 0.7$&$0.94\pm 0.15$ & 0.47&1.36&
$4.0\pm 0.9$& $3.9\pm 0.6$\\ \hline
\end{tabular}
\end{table*}

In the left panel of Fig. 3, we present the measured redshift-space 
two point correlation function in logarithmic intervals
(dots). The clustering is evident, although the correlation
function beyond $r\ge 11 \;h^{-1}$Mpc drops dramatically.
The dashed line corresponds to the best-fitting power law model 
$\xi(r)=(r_{0}/r)^{\gamma}$, determined 
by the standard $\chi^{2}$ minimization procedure, in which each
correlation point is weighted by the inverse of its error.
 
In Fig. 4 we present the iso-$\Delta \chi^{2}$ contours 
(where $\Delta \chi^{2}=\chi^{2}-\chi_{\rm min}^{2}$) in the 
$r_{0}-\gamma$
plane (see the index All).
The $\chi_{\rm min}^{2}$ is the absolute minimum value of 
the $\chi^{2}$. 
In Table 1 we present the best fit parameters, derived 
for $r\ge 1 \;h^{-1}$Mpc in order to avoid 
the signal from the smallest and therefore the highly non-linear 
scales (eg. Vasilyev et al. 2006). Note 
that our results remain robust by varying the upper $r$
limit within the 5 to 20 $h^{-1}$ Mpc range.

\begin{figure}
\mbox{\epsfxsize=8cm \epsffile{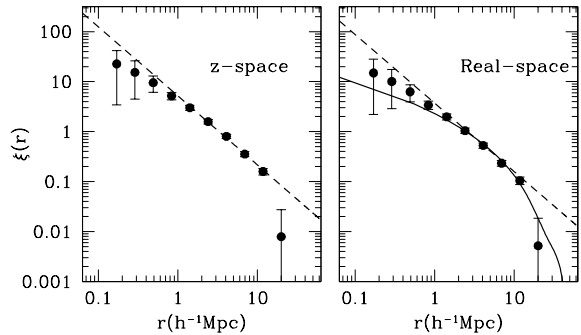}}
\caption{{\em Left panel}: The HIPASS galaxy spatial two-point 
correlation function (points) in redshift space.
The uncertainties were estimated using
100 bootstrap re-samplings of the data (Mo, Jing 
\& B\"orner 1992).   
{\em Right panel}: The HIPASS correlation function in real space.
The dashed line represent the best-fitting power low
$\xi(r)=(r_{0}/r)^{\gamma}$ (see parameters in Table 1), 
while the continuous line represents the best fit $\Lambda$CDM 
($\Omega_{\Lambda}=0.74$) model with 
$b_{0}=0.68$.} 
\end{figure}

The resulting clustering parameters
\footnote{The robustness of our results to 
the fitting procedure was tested using different numbers of bins (spanning from
10 to 20) and we found no significant differences.}
are $r_{0}=3.3 \pm 0.3 h^{-1}$Mpc and 
$\gamma=1.38 \pm 0.24$. Fixing now the correlation function
slope to its nominal value of $\gamma=1.8$, we obtain
$r_{0}= 3.2 \pm 0.2 h^{-1}$Mpc, but now the fit is good for $r\ge 2h^{-1}$Mpc. 
Note that our results are in good agreement with a similar analysis
of the HIPASS data by Ryan-Weber (2006), who find $r_{0}=3.1\pm 0.5h^{-1}$Mpc with 
$\gamma=1.4\pm 0.5$.

It is interesting to compare our results with other recent galaxy
correlation function determinations in the local Universe. In the optical,
Zehavi et al. (2002) and Hawkins et al. (2003) find 
$r_{0}\simeq 5 \;h^{-1}$Mpc and $\gamma\simeq 1.7$ for the SDSS
galaxies, while Norberg et al. (2001) find 
$r_{0}\simeq 4.9h^{-1}$Mpc and $\gamma\simeq 1.7$ for the 2dFGRS galaxies.
In the infrared,
Jing, B\"orner \& Suto (2002) find for the IRAS-PSCz data 
$r_{0}\simeq 3.7h^{-1}$Mpc and $\gamma\simeq 1.7$, while
Gonzalez-Solares et al. (2004) using the ELAIS sources found
$r_{0}\simeq 4.3h^{-1}$Mpc and $\gamma\simeq 2$.
It should be pointed out that the HIPASS galaxy correlation function has a
significantly shallower slope than the optical and infrared galaxy
correlation functions.

\begin{figure}
\mbox{\epsfxsize=8cm \epsffile{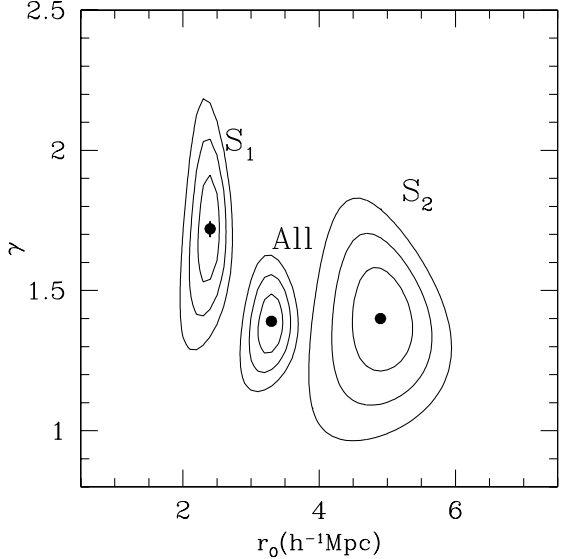}}
\caption{Iso-$\Delta \chi^{2}$ contours  in the $r_{0}-\gamma$ 
parameter space for the whole catalogue and separately for the $S_{1}$
and $S_{2}$ samples. 
The contours correspond to $1\sigma$ ($\Delta \chi^{2}=2.30$), 
$2\sigma$ ($\Delta \chi^{2}=6.17$) and 
$3\sigma$ ($\Delta \chi^{2}=11.8$) uncertainties, respectively.}
\end{figure}

In order to investigate the possible HI mass dependence of the 
correlation function, we divided the sample into two subsamples using
the following mass thresholds: 
(a) $M_{\rm HI}\leq 2.475 \times 10^{9}\;h^{-2}M_{\odot}$, hereafter 
$S_{1}$ sample containing 2389 entries, and 
(b) $M_{\rm HI}> 2.475 \times 10^{9} \;h^{-2}M_{\odot}$, hereafter
$S_{2}$ sample containing 1619 entries.  
We apply our statistical analysis to each HI subsample 
by taking into account the individual redshift selection functions
(see insert of Fig.1). 
The resulting correlation function parameters are presented in
the last two rows of Table 1 and in Fig.4, where we show
the iso-$\Delta \chi^{2}$
contours for both the $S_{1}$ and $S_{2}$ cases, separately.
In Fig. 5, we present the estimated two point redshift correlation 
function for the $S_{1}$ (solid circles) and 
$S_{2}$ (open squares) subsamples, respectively.
The lines correspond to the best-fit power law model.
In particular, for the low HI mass galaxies ($S_{1}$ sample) we obtain
$r_{0}=2.4 \pm 0.3 h^{-1}$Mpc and 
$\gamma=1.72 \pm 0.46$ (continuous line). 
While, for the high 
mass HI galaxies ($S_{2}$ case) we find a 
larger correlation
length, $r_{0}=4.9 \pm 1.0 h^{-1}$Mpc and 
$\gamma=1.40 \pm 0.42$ (see dashed line in Fig. 5).
It is clear that the correlation length increases with 
richness, as expected from its well known richness dependence of
the correlation function (eg. Bahcall \& Burgett 1986).
Note, that in section (2.4) we further investigate 
this issue in more detail.    

\subsection{The HI selected galaxy real-space correlation function}
The previous results are hampered by the fact that the analysis has
been performed in redshift space and therefore
the derived correlation functions are amplified in the linear regime by the
factor $K(\beta)$ (Hamilton 1992, also see Hawkins et al. 2003) given by:
\be
K(\beta)=1+\frac{2\beta}{3}+\frac{\beta^{2}}{5} \;\;,
\ee
where $\beta\simeq \Omega_{\rm m}^{0.6}/b_{0}$, and $b_{0}$ is the
present day bias factor of the HI selected galaxies. Determining this
bias factor and using $\Omega_{\rm m}=0.26$, valid 
for the concordance
cosmological model, we
can evaluate the $K(\beta)$ factor and recover the real-space
HI galaxy correlation function.

To this end, we compare the theoretically expected
mass-tracer correlation function, in the concordance CDM cosmology, with
our derived correlation function.
It is well known (eg. Kaiser 1984; Benson et al. 2000), that according
to the linear biasing ansatz,
the correlation functions of a mass-tracer ($\xi_{\rm obj}$) 
and the dark-matter ($\xi_{\rm DM}$), are related by:
\be
\label{eq:spat}
\xi_{\rm obj}(r)=b_{0}^{2} \xi_{\rm DM}(r) \;\;. 
\ee
We quantify the clustering of the underlying mass distribution 
using the spatial correlation function of the mass, 
$\xi_{\rm DM}(r)$, which 
is the Fourier transform of the spatial power spectrum $P(k)$:
\be
\xi_{\rm DM}(r)=\frac{1}{2\pi^{2}}\int_{0}^{\infty} k^{2}P(k) 
\frac{{\rm sin}(kr)}{kr}{\rm d}k \;\;,
\ee
where $k$ is the comoving wavenumber in units of 
$h$ Mpc$^{-1}$.
As for the power spectrum, we consider that of CDM models: 
$P(k) \approx k^{n}T^{2}(k)$, with
scale-invariant ($n=1$) primeval inflationary fluctuations. 
We utilize the transfer function 
parameterization as in Bardeen et al. (1986), with the approximate
corrections given by Sugiyama's (1995) formula:

$$T(k)=\frac{{\rm ln}(1+2.34q)}{2.34q}[1+3.89q+(16.1q)^{2}+$$
$$(5.46q)^{3}+(6.71q)^{4}]^{-1/4} \;\; .$$
with
\be 
q=\frac{k}{\Omega_{\rm m}h^{2} {\rm exp}[-\Omega_{b}-(2h)^{1/2}
\Omega_{b}/\Omega_{\rm m}]}
\ee
where $\Omega_{b}$ is the baryon density. 
Note that we also use the
non-linear corrections introduced by Peacock \& Dodds (1994).  

In the present analysis we consider the concordance model 
($\Omega_{\rm m}=0.26$) with cosmological parameters, that fit the 
majority of observations, 
$\Omega_{\rm m}+\Omega_{\Lambda}=1$, $H_{0}=100h $km s$^{-1}$ 
Mpc$^{-1}$ with $h\simeq 0.72$ (cf. Freedman et al. 2001;
Peebles and Ratra 2003; Spergel et al. 2003; Tonry et al. 2003; 
Riess et al. 2004; Tegmark et al. 2004; Basilakos \& Plionis 2005; 2006;
Spergel et al. 2006 and references therein), baryonic density 
parameter $\Omega_{\rm b} h^2 \simeq 0.02$ (e.g. 
Olive, Steigman \& Walker 2000; Kirkman et al 2003).
Note that the concordance model 
is normalized to have fluctuation amplitude, in spheres of
8 $h^{-1}$Mpc radius, of $\sigma_{8}\simeq 0.74$
in agreement with both the recent WMAP 3-years results
(Spergel et al. 2006) and the clustering of the XMM X-ray
sources at relatively high redshifts 
(Basilakos \& Plionis 2006). 

In order to derive the HIPASS bias at the present time
we perform a standard $\chi^{2}$ 
minimization procedure between the HIPASS galaxy
correlation function in real space, $\xi_{z}(r)/K(\beta)$, 
and that expected in the concordance cosmological model: 
\be
\chi^{2}(b_{0})=\sum_{i=1}^{n} \left[ \frac{\xi_{z}^{i}(r)/K(\beta) -
\xi_{\rm obj}^{i}(r,b_{0})}
{\sigma^{i}}\right]^{2} \;\;,
\ee 
where $\sigma^{i}$ is the observed correlation function 
uncertainties. The outcome of this analysis provides the HI galaxy
bias factor at the present time: $b_{0}=0.68 \pm 0.10$. 
Thus evidently the HI selected galaxies, in the local Universe, 
are anti-biased with respect to the underlying matter distribution,
implying that they generally trace low-density regions.
In the right panel of Fig. 3, we plot the 
$\xi(r)$ in real space of the HIPASS sample
and the theoretically expected one (continuous line)
for the concordance model using $b_{0}=0.68$.
It is obvious that they compare extremely well.

Performing now the same analysis for the two subsamples, 
we find for the $S_{1}$ subsample (low HI masses) 
a value of: $b_{0}\simeq 0.48$, which again 
means that from the statistical point of view 
the low HI mass galaxies trace low density regions in the local 
universe.
However, for the $S_{2}$ case we get $b_{0}\simeq
0.94$, in agreement with that found using optical galaxies
(eg. Verde et al. 2002; Lahav et al. 2002).

In Table 1 we list our derived values of the HI bias factor, 
$b_{0}$, at the present time, as well as the redshift distortion 
$\beta$ parameter and a measure of the $K(\beta)$ correction. 
Multiplying with $1/K(\beta)$  each bin of
our redshift space correlation function $\xi_{z}(r)$ and 
repeating our fitting procedure 
we derive an estimate of the real-space correlation length 
which is: $r^{\rm re}_{0}=2.4 \pm 0.3 \;h^{-1}$Mpc for $\gamma=1.38$ and
$r^{\rm re}_{0}=2.5 \pm 0.2 \;h^{-1}$Mpc for $\gamma=1.8$.

It is interesting to mention that Ryan-Weber (2006) and Meyer et al. (2007) 
found a somewhat larger correlation length in real space, 
$r^{\rm re}_{0}=3.45 \pm 0.25 h^{-1}$Mpc with 
$\gamma=1.47 \pm 0.08$ and 
$r^{\rm re}_{0}=3.5 \pm 0.7 h^{-1}$Mpc with 
$\gamma=1.9 \pm 0.3$, respectively. This difference should probably be
attributed to the intrinsic uncertainty of the different methods 
used to correct the measured galaxy two-point correlation function 
for redshift-space distortions. 

Note that for the $S_{1}$ and $S_{2}$ subsamples we find 
$r^{\rm re}_{0}=1.8 \pm 0.2 h^{-1}$Mpc and 
$r^{\rm re}_{0}=3.9 \pm 0.6 h^{-1}$Mpc for $\gamma=1.8$.
Finally, we would like to stress that the low-mass HI galaxies appear to have
the lowest correlation length among all extragalactic populations
studied to-date.

\begin{figure}
\mbox{\epsfxsize=9cm \epsffile{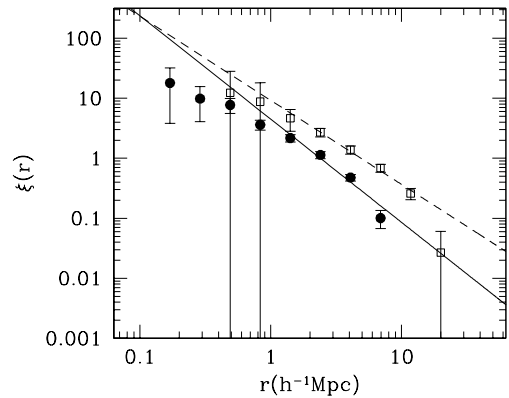}}
\caption{The spatial two-point correlation function in redshift space
for the $\rm S_{1}$ (solid points) and 
$\rm S_{2}$ (open squares) subsamples. 
The error bars are estimated using the bootstrap procedure. 
The lines represent the best-fit power low model,
$\xi(r)=(r_{0}/r)^{\gamma}$ (see parameters in Table
1).} 
\end{figure}

\subsection{The Environment of HI galaxies}
In order to investigate the cause for the relatively low-amplitude of the
HI galaxy correlation function, we decided to correlate the positions
of the HI galaxies with the reconstructed smoothed density field of the PSCz
flux limited (0.6 Jy) IRAS galaxy catalogue. We use the reconstructed density
field of Branchini et al. (1999), smoothed as in Plionis \& Basilakos
(2001) utilizing a Gaussian kernel on a 3D grid with grid-cell size
and a smoothing scale ($R_{{\rm sm}}$) equal to $5 \; h^{-1}$ Mpc. We
also correct for effects related to the unavoidable convolution of 
the constant smoothing window with the degradation of the redshift 
selection function (see Plionis \& Basilakos 2001 for details).
We then cross-correlate the positions of each HI galaxy with the PSCz
smoothed density field and assign to each the overdensity of its
nearest grid cell. In this way we create the 1-point PSCz overdensity
probability distribution which is related to the HI galaxy positions
and compare it with the corresponding full PSCz 1-point distribution function.
Note that we exclude from our analysis the inner volume ($r< 15\;
h^{-1}$ Mpc) in order to avoid discreteness effects, ie., associating
a large number of HIPASS galaxies with only a few PSCz density
grid-cells (16\% of HI galaxies with 0.33\% of the total grid-cells).

\begin{figure}
\mbox{\epsfxsize=8cm \epsffile{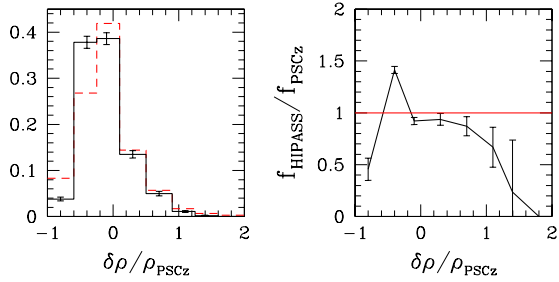}}
\caption{{\em Left Panel:} The 1-point PSCz overdensity distribution
function related to the positions of the HI galaxies (solid line) and
the corresponding full PSCz distribution (dashed line). {\em Right
Panel:} The ratio of the two distributions.}
\end{figure}

In the left panel of Fig. 6 we present the two frequency
distributions, for which the Kolmogorov-Smirnov two-sided test shows that
the probability of them having the same parent distribution is
$<10^{-10}$. In the right panel of the same figure we show the ratio
of the two distributions as a function of $\delta \rho/\rho$.
If the HI galaxies would trace the same structures as the IRAS-PSCz
galaxies then the two distribution should be indistinguishable and the
ratio would be $\sim 1$. The results presented in this panel are very
informative regarding the local environment of HI galaxies. We see
that the HI galaxies avoid high density regions, as well as the very
low density regions. They prefer, with respect to IRAS galaxies,
regions of quite low galaxy density ($-0.4 \mincir
\delta\rho/\rho_{\rm PSCz} \mincir 0$).

In order to verify the other outcome of our HI galaxy correlation function
analysis, ie., the fact that there appears to be a dichotomy between
the clustering properties of the low and high mass subsamples of HI
galaxies, we decided to compare the PSCz density distribution
corresponding to the positions of the high and low HI mass galaxies.
In Fig. 7 we show this ratio and it is evident that with respect to
the high mass HI galaxies, the low mass ones populate typically only 
under-dense
void-like regions ($\delta\rho/\rho_{\rm PSCz} \mincir 0$). {\em We find that
75\% of low-mass HI galaxies are associated with negative PSCz
overdensities, with 40\% having $\delta\rho/\rho_{\rm PSCz} <-0.3$.} 
This fact corroborates our previous clustering analysis 
and the resulting extremely low-bias parameter of these 
HI galaxies ($b_{0}\simeq 0.48$).
It should also be kept in mind that with respect to the 
optical, infrared galaxies typically avoid the 
high-density regions, like those related to clusters of galaxies,
a fact which translates to a PSCz correlation length 
of $r_{0}\simeq 3.7h^{-1}$Mpc (Jing et al. 2002), while
the PSCz bias parameter, within the concordance
cosmological model ($\Omega_{\rm m}=0.26$), has been found to be 
$b_{\rm PSCz} \simeq 0.9$ (eg., Basilakos \& Plionis 2006).

Furthermore, these results are very interesting in themselves because they
indicate that low-density void-like regions could well be filled by
such HI low-mass
galaxies, although it also appears that the lowest density regions are
avoided also by such galaxies. We believe that this issue will be
settled in the next few years with the new {\sc alfalfa}
survey, which is expected to observe about 16000 HI galaxies,
most of them of low mass, in 7000 square degrees 
(Giovanelli et al. 2005a; 2005b).


\begin{figure}
\mbox{\epsfxsize=7cm \epsffile{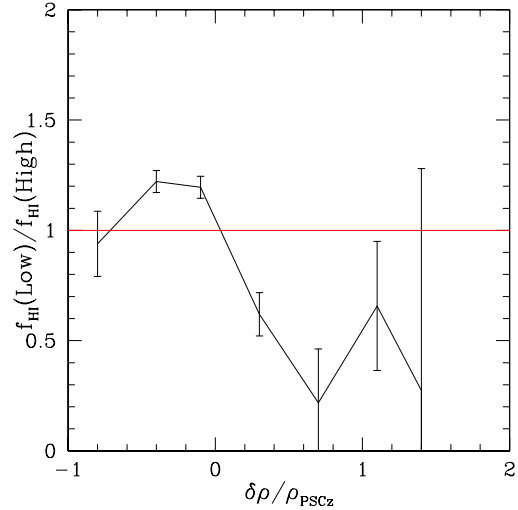}}
\caption{The ratio of the 1-point PSCz overdensity distribution
functions of the low and high HI mass galaxies.}
\end{figure}

\section{HI Galaxies Bias Evolution}
In order to understand better the importance of the measured HI galaxy
clustering, we investigate the evolution of 
the HI bias factor as a function of redshift.
The evolution of neutral hydrogen as a function of redshift
is a powerful cosmological tool because it indicates the rate of 
transformation of gas into stars
and thus the gas consumption and the evolution of star formation
in the Universe.

Over the past two decades, based on different assumptions, a number 
of bias evolution models have been
proposed (eg.  Nusser \& Davis 1994; Fry 1996; Mo \& White 1996;
Matarrese et al. 1997; Tegmark \& Peebles 1998; 
Basilakos \& Plionis 2001). Here we will use only models that
have been shown to describe relatively well the evolution of bias
even beyond $z\sim1$. These are:

\begin{itemize}

\item {\em Merging Bias Model} (hereafter B1): 
Mo \& White (1996) have developed a model for 
the evolution of the the so-called correlation bias, 
using the Press-Schechter formalism.
Utilizing a similar formalism, Matarrese et al. (1997) extended 
the Mo \& White (1996)
results to include the effects of different mass scales (see also 
Bagla 1998). 
In this case the expression which describes the bias evolution is
\be
b_{\rm B1}(z)=0.41+\frac{(b_{0} - 0.41)}{D^{\beta}(z)} \;\;\; ,
\ee
with $\beta \simeq 1.8$.
Note that $D(z)$ is the linear growth rate of clustering (cf. Peebles 1993)
\footnote{$D(z)=(1+z)^{-1}$ for an Einstein-de Sitter Universe.} 
scaled to unity at the present time.

\be\label{eq:24}
D(z)=\frac{5\Omega_{\rm m} E(z)}{2}\int^{\infty}_{z} \frac{(1+x)}{E^{3}(x)} 
{\rm d}x\;\;. 
\ee
with 
\be\label{eq:4}
E(z)=\left[ \Omega_{\rm m}(1+z)^{3}+\Omega_{\Lambda} \right]^{1/2} \;\;.
\ee

\item {\em Bias from Linear Perturbation Theory} (hereafter B2): 
Basilakos \& Plionis (2001, 2003), using linear
perturbation theory and the 
Friedmann-Lemaitre solutions of the cosmological
field equations have derived analytically the functional form
for the evolution of the 
linear bias factor, $b$, between the background matter and 
a mass-tracer fluctuation field. The main assumptions of this model is
that the DM and galaxies share the same velocity field and that the mass
tracer population is conserved in time.
For the case of a spatially flat $\Lambda$ cosmological model
($\Omega_{\rm m}+\Omega_{\Lambda}=1$), the bias 
evolution can be written as:

\be\label{eq:84}
b_{\rm B2}(z)={\cal C}_{1} E(z)+{\cal C}_{2}E(z)I(z)+1
\ee
with
\be
I(z)=\int_{1+z}^{\infty} \frac{y^{3} {\rm d}y}
{[\Omega_{\rm m} y^{3}+\Omega_{\Lambda}]^{3/2}} \;\;.
\ee
Note that this model gives a family of bias
curves, due to the fact that it has two unknown parameters, 
(the integration constants ${\cal C}_{1},{\cal C}_{2}$). 
Basilakos \& Plionis (2001, 2003) 
compared the B2 bias evolution model with other models as well
as with the HDF (Hubble Deep Field) biasing 
results (Arnouts et al. 2002), and found a good consistency.
Of course in order to obtain partial solutions for $b(z)$ we need 
to estimate the 
values of the constants ${\cal C}_{1}$ and ${\cal C}_{2}$, which means that
we need to calibrate the $b(z)$ relation using two different epochs:
$b(0)=b_{0}$ and $b(z_{1})=b_{1}$. 

Therefore, utilized the general bias solution (see eq.13), 
it is routine to obtain the expressions for the above 
constants as a function of $b_{0}$ and $b_{1}$:
\be
{\cal C}_{1}=\frac{ (b_{0}-1)E(z_{1})I(z_{1})-(b_{1}-1)E(0)I(0)}
{E(0)E(z_{1})[I(z_{1})-I(0)]}
\ee
\be
{\cal C}_{2}=\frac{ E(0)(b_{1}-1)-E(z_{1})(b_{0}-1)}
{E(0)E(z_{1})[I(z_{1})-I(0)]} \;\;\;,
\ee
where for the present epoch we have: $b_{0}\simeq 0.68$, $E(0)=1$ and 
$I(0)\simeq 11.54$. 

\subsection{The HI bias factor in the distant Universe}
The hyperfine transition responsible for the HI emission is very weak
and therefore with current telescopes, studies of the emission at 21-cm 
are limited to the very local Universe ($z\le 0.2$). This implies that
the estimation of the bias parameter $b_{1}$ 
of HI sources in the distant universe ($z>1.5$) 
would not be an easy task. At such redshifts, our knowledge on the 
neutral gas comes mostly from the absorption
lines detected in the spectra of background quasars. 

At redshifts $1.5\le z \le 5.0$ 
most of the cosmic HI mass is contained in Damped Ly$\alpha$
systems (DLAs, e.g.  Wolfe et al. 1986, Lanzetta,  Wolfe \& Turnshek
1995).  The DLAs  are quasar  absorption line  systems with  HI column
density $N_{HI} \ge 2 \times 10^{20}$ atoms cm$^{-2}$ 
(see Peroux et al. 2001; Wolfe et al. 2005) and they are composed of 
predominantly neutral gas.  
Wolfe (1986) established  the given column density  threshold  so
that it corresponds to the surface density limit of the 
local 21-cm observations at that time, 
which roughly corresponded to the transition from primarily
ionized gas to predominantly neutral gas (e.g., Viegas 1995, 
Prochaska 1999, Vladilo et al. 2001).
Therefore, these systems play a vital role in the
structure formation because the large column 
densities can protect large amounts of neutral gas from the ionizing
background (Zwaan \& Prochaska 2006) 
and thus produce a favorable environment for star formation.  
To this end it is interesting to mention that 
Zwaan et al. (2005b) have found that
in the redshift range of $z\sim 5$ to $z=0$ the 
vast majority of the mass density in HI sources is locked up in column 
densities of $N_{\rm HI} > 2\times 10^{20}$atom cm${^2}$. 

We will be using the above ideas to calibrate the B2-model,
based on the Cooke et al. (2006) value of
the bias for Damped Lyman-$\alpha$ systems which is $b(3) \simeq 2.4$ 
(see solid point in Fig.8 )
\footnote{For $z=3$ and $\Omega_{\rm m}=1-\Omega_{\Lambda}=0.26$ 
we get $E(3)\simeq 4.17$ and $I(3)\simeq 7.47$}. 

\end{itemize}

\begin{figure}
\mbox{\epsfxsize=8.3cm \epsffile{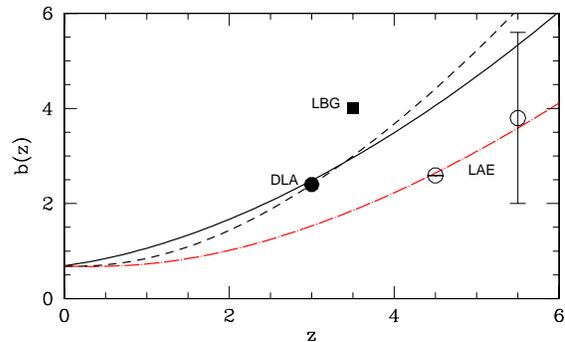}}
\caption{The evolution of the HI galaxy bias factor for
the B1 (continuous line) and B2 (dashed line) bias models, respectively.
The solid circle represents the observational bias for the 
Damped Lyman-$\alpha$ systems (DLA) of Cooke et al. (2006). The square-like
symbol corresponds to Lyman break galaxies (LBG, Steidel et al. 1998)
while the open circle represents the bias for the Lyman-$\alpha$ emitters (LAE)
derived by Ouchi et al. (2005) and 
Kova\v{c} et al. (in preparation). Note that
the dot-dashed line represents the bias behavior for the LAE galaxies.}
\end{figure}

\subsection{The global HI bias results}
Knowning the HI galaxy bias factor at the present time, we can use
eqs.(15) and (16) and the observed DLA bias factor at $z\simeq 3$ 
to plot the B2 bias evolution model 
for the concordance cosmological model 
(see Fig.8). We also plot the B1 evolution
model, which is uniquely determined from the present day HI galaxy 
bias factor (B1 continuous line and B2 dashed line). 
Evidently, up to $z\le 1.3$
we have an anti-bias picture but 
due to the fact that the bias is a monotonically 
increasing function of redshift, 
for both B1 and B2 biasing models, the HI sources at higher redshifts
are biased with respect to the underlying matter distribution.  
Note also that 
both models provide almost the same 
bias evolution behavior, while for $z\magcir 5$ the two models
diverge.

We further compare our analytic solutions 
with observations of: (a) Lyman-break galaxies (hereafter LBG; see 
solid square in Fig.8) which are strongly biased with respect 
to the underlying mass and have $b(3.4) \simeq 4$ (Steidel et al. 1998;
Adelberger et al. 1998; Kashikawa et al. 2006), and
(b) Lyman-$\alpha$ emitters (hereafter LAE; see open circles in Fig.8) with
$b(5.7)=3.8\pm 1.8$ (Ouchi et al. 2005) and $b(4.5)\simeq 2.6$
(Kova\v{c} et al. in preparation). For the LAE case we
present in Fig.8 the B2 bias evolution model,
anchored at $z=4.5$ (dot dashed line)\footnote{For $z=4.5$ we 
get $E(4.5)\simeq 6.64$ and $I(4.5)\simeq 6.41$}.
Within this bias evolution model the LAE galaxies appear anti-biased
up to $z\simeq 2.1$, while beyond a redshift of $z\sim 3$ they become
strongly biased with respect to the underlying mass fluctuations.

The strong biasing predictions, in agreement 
with those found from simulations of galaxy formation
(eg. Kauffmann et al. 1999), imply that Lyman-$\alpha$ systems, 
at high redshifts, 
are formed at the highest peaks of matter density field (eg. Mo \& White 2002).
The last few years many authors
using Lyman-$\alpha$ systems have found high overdensities ($\delta >50$) at 
high redshifts. Indeed, Steidel et al. (1998)
using LBGs found a protocluster at
$z\simeq 3.1$. Recently, Venemans et al. (2004), 
Miley et al. (2004), Ouchi et al. (2005) and Intema et al. (2006)
found protoclusters of LAE galaxies at high redshifts ($z\simeq 4$).
Also a cross correlation analysis 
has shown (Cooke et al. 2006) that the LBGs are associated with the 
DLA systems at the same redshift, a fact which implies that
perhaps the LBGs trace the same parent system as the DLAs (Cooke et al. 2005).

\section{Conclusions}
We have studied the clustering properties of the 
HI 21-cm emission line sources from HIPASS catalogue 
in redshift space. 
Modeling two point correlation function
as a power law, $\xi(r)=(r_{0}/r)^{\gamma}$, we find 
$r_{0}=3.3 \pm 0.3 h^{-1}$ Mpc and
$\gamma=1.38 \pm 0.24$.
Fixing the slope to its universal value $\gamma=1.8$, we obtain
$r_{0}=3.2\pm 0.2 h^{-1}$Mpc.

Comparing the measured spatial correlation function  
for the HI selected galaxies
with the theoretical predictions of the preferred 
$\Lambda$CDM cosmological model 
($\Omega_{\rm m}=1-\Omega_{\Lambda}=0.26$) and a 
bias evolution model, we find that the present bias value is 
$b_{0}\simeq 0.68$, suggesting that the HI selected galaxies
are located in relatively underdense regions 
in the local Universe, in agreement with previous studies 
(Grogin \& Geller 1998). Using the derived value of the bias we
determine the redshift-space distortion parameter, $K(\beta)$, and 
use it to derive the real-space correlation function, for which we have:
$r^{\rm re}_{0}=2.4 \pm 0.3h^{-1}$Mpc with $\gamma=1.38$
and $r^{\rm re}_{0}=2.5 \pm 0.2h^{-1}$Mpc for $\gamma=1.8$.

Dividing our HI galaxy sample into two richness subsamples we find 
(a) for $M_{\rm HI}\leq 2.475 \times 10^{9}h^{-2}M_{\odot}$ that
$r^{\rm re}_{0}\simeq 1.8 h^{-1}$ Mpc 
(for $\gamma=1.8$) and $b_{0}\simeq 0.48$ 
while (b) for $M_{\rm HI}> 2.475 \times 
10^{9}h^{-2}M_{\odot}$ that  
$r^{\rm re}_{0}\simeq 3.9 h^{-1}$ Mpc 
(for $\gamma=1.8$) and $b_{0}\simeq 0.94$. 
Therefore, 
the low mass HI galaxies should trace  
low densities, due to very low biasing found while
the high mass HI galaxies should trace average galaxy densities
as optical galaxies do ($b_{0}\simeq 1$).
Indeed, we corroborate these results by correlating the HI galaxy
positions with the IRAS-PSCz density field, smoothed over scales of 5
$h^{-1}$ Mpc, to find that indeed the HI galaxies are typically 
found in negative overdensity regions 
($\delta\rho/\rho_{\rm PSCz} \mincir 0$), even more so the
low-mass HI galaxies. We conclude that such
galaxies could be the typical population of galaxies in
void-like regions.

Finally, we investigate the redshift evolution of 
the HI galaxy linear bias factor and we find that the anti-bias 
behavior extends up to $z\le 1.3$. 

\section* {Acknowledgements}
We would like to thank the anonymous referee for his/her useful 
comments and suggestions.
Dr. Kova\v{c} now is a post-doc at the 
Institut f\"{u}r Astronomie, ETH in Z\"{u}rich.

{\it In memory of Nikos Voglis}. This paper is 
dedicated to Professor Nikos Voglis, who suddently
passed away at the age of 58, in his office at the Academy of Athens,
while working on this manuscript. His colleagues and students
will miss his exceptional intellect and enthusiasm.


{\small
\begin{thebibliography}{}
\bibitem[]{} Adelberger, K. L., Steidel, C. C., Giavalisko, M.,
  Dickinson M., Pettini, M., \&, Kellogg, M., 1998, ApJ, 505, 18
\bibitem[]{} Arnouts, S., et al., 2002, MNRAS, 329, 355
\bibitem[]{} Bagla, J. S., 1998, MNRAS, 299, 424
\bibitem[]{} Bahcall, N., \&, Burgett, W. S., 1986, ApJ, 300, L35
\bibitem[]{}Barnes et al., 2001, MNRAS, 322, 486
\bibitem[]{}Bardeen, J.M., Bond, J.R., 
Kaiser, N. \& Szalay, A.S., 1986, ApJ, 304, 15
\bibitem[]{} Basilakos, S. \& Plionis, M., 2001, ApJ, 550, 522
\bibitem[]{} Basilakos, S. \& Plionis, M., 2003, ApJ, 593, L61
\bibitem[]{} Basilakos, S. \& Plionis, M., 2005, MNRAS, 360, L35
\bibitem[]{} Basilakos, S. \& Plionis, M., 2006, ApJ, 650, L1
\bibitem[]{} Basilakos, S. \& Plionis, M., 2006, MNRAS, 373, 1112
\bibitem[]{}Benson A. J., Cole S., Frenk S. C., 
Baugh M. C., Lacey G. C., 2000, MNRAS, 311, 793
\bibitem[]{}Branchini, E., et al., 1999, MNRAS, 308, 1
\bibitem[]{}Coil, A. L., Newman, J. A., Cooper, M. C., Davis, M., Faber, S. M.,
Koo, D. C., Willmer C. N. A, 2006, ApJ, 644, 671
\bibitem[]{}Colless, M., et al., 2001, MNRAS, 328, 1039
\bibitem[]{}Cooke, J. W., Wolfe, A. M., Prochaska, J. X., Gawiser, E.,
2005, ApJ, 621, 596
\bibitem[]{}Cooke, J. W., Wolfe, A. M., Gawiser, E., Prochaska, J. X.,
2006, ApJ, 636, L9
\bibitem[]{}Davis, M., et al., 2003, Proc. SPIE, 4834, 161, 
{\it astro-ph/0209419}  
\bibitem[]{}Doyle, M. T., et al., 2005, MNRAS, 361, 34
\bibitem[]{} Efstathiou, G., Bernstein, G., Katz, N., Tyson, J. A., 
Guhathakurta, P., 1991, ApJ, 380, L47
\bibitem[]{}Freedman, W., L., et al., 2001, ApJ, 553, 47
\bibitem[]{}Fry J.N., 1996, ApJ, 461, 65
\bibitem[]{} Giovanelli, R., et al., 2005a, AJ, 130, 2598
\bibitem[]{} Giovanelli, R., et al., 2005b, AJ, 130, 2613
\bibitem[]{}Gonzalez-Solares, E. A., et al., 2004, MNRAS, 352, 44
\bibitem[]{}Grogin, N. A., \&, Geller, M. J., 1998, ApJ, 505, 506
\bibitem[]{}Guzzo et al., 2007, XXVIth Astrophysics Moriond Meeting: "From Dark Halos to Light", March 2006, 
proc. edited by L.Tresse, 
S. Maurogordato and J. Tran Thanh Van (Editions Frontieres), {\it astro-ph/0701273}  
\bibitem[]{}Jing, Y. P., B$\ddot {\rm o}$rner, G, Suto, Y., 
2002, ApJ, 564, 15
\bibitem[]{}Hamilton, A. J. S., 1992, ApJ, 385, L5
\bibitem[]{}Hawkins, Ed, et al., 2003, MNRAS, 346, 78
\bibitem[]{}Intema, H. T., Venemans, B. P., Kurk, J. D., Ouchi M., 
Kodama, T., H. Rottgering J. A., Miley, G. K., Overzier, R. A., 2006,
A\&A, 456, 433
\bibitem[]{}Kaiser N., 1984, ApJ, 284, L9
\bibitem[]{}Kashikawa, N., et al., 2006, ApJ, 637, 631
\bibitem[]{} Kauffmann G., Golberg, J. M., Diaferio A., White S. D. M., 
1999, MNRAS, 307, 529
\bibitem[]{}Kirkman, D., Tytler, D., Suzuki, N., O'Meara, J.M., Lubin,
D., 2003, ApJS, 149, 1
\bibitem[]{}Lahav, O., et al., 2002, MNRAS, 333, 961
\bibitem[]{}Lanzetta, K. M., Wolfe, A. M., Turnshek, D. A., 1995, ApJ,
  440, 435
\bibitem[]{}Le F\'{e}vre, et al. 2005, A\&A, 439, 845
\bibitem[]{} Mataresse, S., Coles, P., Lucchin, F., Moscardini, L.,
  1997, MNRAS, 286, 115
\bibitem[]{}Meyer, M. J., et al., 2004, MNRAS, 350, 1195
\bibitem[]{}Meyer, M. J., Zwaan, M. A., Webster, R. L., Brown, M. J. I., 
Staveley-Smith, L., 2007, ApJ, 654, 702
\bibitem[]{}Miley, G. K.., et al., 2004, Nature, 427, 47
\bibitem[]{}Mo, H. J., Jing, Y. P., 
B$\ddot {\rm o}$rner, G., 1992, ApJ, 392, 452 
\bibitem[]{}Mo, H.J, \& White, S.D.M  1996, MNRAS, 282, 347
\bibitem[]{}Mo, H.J, \& White, S.D.M  2002, MNRAS, 336, 112
\bibitem[]{}Norberg, P., et al., 2001, MNRAS, 328, 64
\bibitem[]{}Nusser, M., \& Davis, M., 1994, ApJ, 421, L1
\bibitem[]{}Olive, K.A., Steigman, G., Walker, T.P., 2000, 
Phys.Rep., 333, 389
\bibitem[]{}Oliver, S., et al., 2000, MNRAS, 316, 749
\bibitem[]{}Ouchi, M., et al., 2005, ApJ, 620, l10
\bibitem[]{}Peacock, A. J., \&, Dodds, S. J., 1994, MNRAS, 267, 1020
\bibitem[]{}Peebles P.J.E., 1993, Principles of Physical Cosmology, 
Princeton University Press, Princeton New Jersey
\bibitem[]{}Peebles P.J.E., \&, Ratra, B., 2003, RvMP, 75, 559
\bibitem[]{}Peroux, C., Storrie-Lombardi, L. J., McMahon, R. G., 
Irwin, M., \&, Hook, I. M., 2001, AJ, 121, 1799
\bibitem[]{}Plionis, M., \&, Basilakos, S., 2001, MNRAS, 327, L32
\bibitem[]{}Prochaska, J. X., 2001, ApJ, 511, L71
\bibitem[]{} Riess, A. G., et al., 2004, ApJ, 607, 665
\bibitem[]{}Ryan-Weber, E. V., MNRAS, 2006, 367, 1251
\bibitem[]{}Saunders, W., et al., 2000, MNRAS, 317, 55
\bibitem[]{}Spergel, D. N., et al., 2003, ApJS, 148, 175
\bibitem[]{}Spergel, D. N., et al., 2007, ApJ, in press, 
{\em astro-ph/0603449}
\bibitem[]{}Steidel C.C., Adelberger L.K., Dickinson M., 
Giavalisko M., Pettini M., Kellogg M., 1998, ApJ, 492, 428
\bibitem[]{}Sugiyama, N., 1995, ApJS, 100, 281
\bibitem[]{}Tegmark M. \& Peebles P.J.E, 1998, ApJL, 500, L79
\bibitem[]{}Tegmark M., et al., 2004, Phys. Rev. D., 69, 3501
\bibitem[]{}Tonry, et al. , 2003, ApJ, 594, 1
\bibitem[]{}Wolfe, A. M., Tumshek, D. A., Smith, H. E.,
Cohen, R. D., 1986, ApJS, 61, 249
\bibitem[]{}Wolfe, A. M., Gawiser, E., Prochaska, J. X., 2005, ARA\&A, 43, 861
\bibitem[]{}Wong, O. I., et al., 2006, MNRAS, 371, 1855
\bibitem[]{}Vasilyev, N. L., Baryshev, Yu. V., 
Sylos Labini, F., 2006, A\&A, 447, 431
\bibitem[]{}Venemans, B. P., et al., 2004, A\&A, 427, L17
\bibitem[]{}Verde, L., 2002, MNRAS, 335, 432
\bibitem[]{}Viegas, S. M., 1995, MNRAS, 276, 268
\bibitem[]{}Vladilo, G., Centurion, M., Bonifacio, P., Howk, J. C.,
  2001, 557, 1007
\bibitem[]{}York, D. G., 2000, AJ, 120, 1579
\bibitem[]{}Zehavi I., et al., 2002, ApJ, 571, 172
\bibitem[]{}Zwaan, M. A., et al., 2004, MNRAS, 350, 1210
\bibitem[]{}Zwaan, M. A., Meyer, M. J., Staveley-Smith, L., Webster, R. L.,
2005, MNRAS, 359, L30
\bibitem[]{}Zwaan, M. A., van der Hulst, J.M ., Briggs H. F.,
  Verheijen M. A. W., Ryan-Weber, E. V., MNRAS, 2005b, 364, 1467
\bibitem[]{}Zwaan, M. A., \&, Prochaska, J. X., 2006, ApJ, 643, 675

\end{thebibliography}
}
\end{document}